# Superfluid density as a guide to optimal superconductivity in doped low dimensional antiferromagnets

Christos Panagopoulos *

Department of Physics, University of Crete, and Foundation for Research and Technology-Hellas, P.O.Box 2208, 71003 Heraklion, Crete, Greece

**Abstract**

Following the direct observation of abrupt changes in the superconducting ground state in doped low dimensional antiferromagnets, we have identified a phase transition where superconductivity is optimal. The experiments indicate the presence of a putative quantum critical point associated with the emergence of a glassy state. This mechanism is argued to be an intrinsic property and as such largely independent of material quality and the level of disorder.

* On leave from the Department of Physics, University of Cambridge, Cambridge CB3 0HE, United Kingdom

**Introduction**

It is a privilege to be invited to contribute to this volume in honour of Professor David Shoenberg with whom I had the opportunity to discuss and receive advice several times since my arrival as a graduate student in Cambridge in 1994. A number of my works have been based on what I view as merely an extension to a clever experimental technique developed by David approximately half a century ago. David introduced a novel method to determine the magnetic penetration depth of isotropic superconductors [1]. With some modifications we performed, this technique is now applicable to highly anisotropic superconducting materials. This "extension" formed the base for a number of my works on the very complex subject of high temperature superconductivity [2-12]; in particular, the study of an electromagnetic ground state which in a given region and under experimentally controlled conditions may change from an insulator to a metal and even a superconductor.

High temperature superconductors (HTS) are an excellent example of low dimensional magnets with a well controlled tuneable ground state [13-15]. These materials allow the study of the relationship between magnetism and unconventional superconductivity in low dimensional antiferromagnets, a fundamental problem in modern condensed matter physics. The parent compounds of HTS are antiferromagnetic insulators with weak inter-plane interaction that become superconducting upon charge doping. In the prototype $La_2CuO_4$ for example, the Néel temperature decreases fast with doping by replacing $La^{3+}$ with $Sr^{2+}$ and at the characteristic value $x_{sc}$=0.055 holes/planar-Cu-atom the system undergoes an insulator-superfluid transition. The emergence of bulk superconductivity is preceded with unusual non-Fermi liquid behaviour above the superconducting transition temperature $T_c$. With further doping, the material seems to progress to a Fermi-liquid state, with the formation of a coherent 3D Fermi surface and the eventual demise of superconductivity in the neighbourhood of $x \sim 0.27$.

Because HTS are close to a phase transition they can also exhibit large responses to small external signals. These signals can drive the system from one phase to the other, dramatically changing the properties of the material. By changing the parameters of the system we may tune it to be exactly at or on the cusp of the transition, where electrons behave most collectively. In the case of pure 3D systems like some of the f-electron compounds for example, there have been even suggestions for the presence of a quantum critical point with an antiferromagnetic order vanishing at optimal superconductivity [16]. In the quasi-2D HTS however, an actual long range order may be replaced by an order associated with glassiness. For an experimentalist, probing the glassy phase and in particular searching for the possible presence of quantum critical behaviour through changes in the ground state is a difficult task and the outcome of the study such dynamic processes would depend on the probe used. Hence identifying the right tool(s) is of primary importance.

The superfluid density, perhaps the most fundamental quantity of superconductivity, is a property which allows us probe directly changes in the superconducting ground state of these low dimensional complex systems. In the middle of last century, Shoenberg proposed and tested a method to measure the magnetic penetration depth (inversely proportional to the superfluid density) of isotropic conventional superconductors [1]. We worked for several years in developing an "extension" to this method to study the HTS. The advantage of the Shoenberg-technique is that once optimised for the system under investigation it can be applied to almost any newly discovered superconductor and with high accuracy. We may determine accurately both the absolute value and temperature dependence of the superfluid density. Consequently we have been able to make a number of discoveries [2-12]. Among the discoveries, and subject of this article, was the identification of well-defined changes in the superconducting ground state of HTS [8,11,12]. In fact to the best of my knowledge these may be the only changes in the superconducting ground state of a material measured by tuning its ground state using a single parameter while measuring a fundamental property of superconductivity. Motivated by our observations we employed techniques investigating the magnetic ground state to search for possible changes accompanying those of superconductivity [11]. We discovered simultaneous changes in both the magnetic and superconducting ground states with a novel quantum glass transition - a quantum critical point intercepted by glassiness, present where superconductivity is optimal.

**Experimental**

Here I will discuss results obtained only on one set of samples. I will concentrate on the archetypal HTS family, $La_{2-x}Sr_xCuO_4$ (La-214), which allows one to span the whole phase diagram of these materials [13-15]. Although the particular family might be considered somewhat more susceptible to disorder than other members of the HTS-group, it serves very well our purpose of studying the low energy dynamics as the latter would be enhanced here and therefore subtle features not measurable otherwise may now become identifiable. Except the single crystals used for the thermodynamic studies, we have also synthesised polycrystalline samples of $La_{2-x}Sr_xCu_{1-y}Zn_yO_4$ ($x$=0.03-0.24 and $y$=0, 0.01 and 0.02) using solid-state reaction followed by quenching and subsequent oxygenation. We note that in La-214, $x$ is the carrier concentration per $CuO_2$ plane. The samples were characterised by powder x-ray diffraction, micro-Raman spectroscopy as well as extensive transport and thermodynamic measurements and found to be phase pure. Their values of the superconducting transition temperature, $T_c$, agree with published data for powders and single crystals [17,18]. The lattice parameters were also in good agreement with published data, where available [17]. Care was taken to ensure high purity and homogeneity. For example, we dry all precursor powders; we react, grind, mill, re-press and re-sinter at least 4 times to ensure extrinsic homogeneity. In the Zn-doped samples we encapsulate to ensure minimal Zn loss. To avoid possible phase separation we quench from the last high temperature reaction then begin our oxygen anneal at 750°C and apply a slow cool over many days to 350°C to ensure full oxygenation of vacant oxygen sites associated with Sr substitution. Micro-analytical spectroscopic studies show no signature of impurities or inhomogeneity.

To study the evolution of the ground state we measured the superfluid density. The superfluid density, $\rho_s \sim \lambda_{ab}^{-2}$, was determined from measurements of the in-plane magnetic penetration depth $\lambda_{ab}$ using the low-field $ac$-susceptibility technique for grain-aligned powders. To magnetically align the samples, the powder was first lightly ground by hand (in argon atmosphere [19]) to remove aggregates and passed through sieves. In some materials it is common to obtain small grain agglomerates of the order of a few microns. Randomly oriented grain agglomerates can be a cause of poor alignment, and to eliminate these, powders obtained after sieving were ball milled in ethanol and dried after adding a defloculant. Scanning electron microscopy confirmed the absence of grain boundaries. The powders were mixed with a 5-min curing epoxy and aligned in a static field of 12 T at room temperature. Debye-Scherrer x-ray scans showed that more than 90%, often as high as 95%, of the grains had their $CuO_2$ planes aligned to within approximately 1.8, or smaller, degrees [20]. The *ac*-susceptibility measurements were performed down to 1.2K with a home-made susceptometer using miniature coils. The Earth's field was screened out using $\mu$-metal shield. Measurements were performed for $H_{ac}$=1Gauss rms at $f$=333Hz. The separation of the grains and the absence of weak links were confirmed by checking the linearity of the signal for $H_{ac}$ from 0.3 to 3Gauss rms and $f$ from 33 to 333Hz. The data were analysed using the extended-Shoenberg method [2-12], and the absolute values and variation of $\lambda_{ab}$ with temperature were obtained. Furthermore, the values of $\lambda_{ab}^{-2}(0)$ for some of the samples were also confirmed by standard transverse field $\mu$SR experiments [8,11].

A technique extensively used to study the low energy ($10^{-9} - 10^{-6}$ s) dynamics in magnetic materials and alloys is muon spin relaxation ($\mu$SR) [21]. It has been successfully employed in the study of HTS and provided some of the earliest evidence for the presence of low energy spin excitations in these materials [22-27]. It is therefore instructive to perform a systematic $\mu$SR investigation of the low-energy spin dynamics across the *T-x* phase diagram in several HTS and preferably in small increments in doping state. Our $\mu$SR studies were performed at the pulsed muon source, ISIS Facility, Rutherford Appleton Laboratory. Spectra were collected down to as low as 40mK thus allowing the temperature dependence of slow spin fluctuations to be studied to high doping. In a $\mu$SR experiment, 100% spin-polarised positive muons implanted into a specimen precess in their local magnetic environment. Random spin fluctuations will depolarise the muons provided they do not fluctuate much faster than the muon precession. The muon decays with a life-time of 2.2$\mu$s, emitting a positron preferentially in the direction of the muon spin at the time of decay. By accumulating time histograms of such positrons one may deduce the muon depolarisation rate as a function of time after implantation. The muon is expected to reside at the most electronegative site of the lattice, in this case 1Å away from the apical $O^{2-}$ site above and below the $CuO_2$ plane, so the results reported here are dominated by the magnetic correlations in the $CuO_2$ planes [26].

**Results and Discussion**

We have obtained systematic results on the effects of carrier concentration on the superfluid density of La-214. For $x>x_{opt} \sim 0.19$ (where $\rho_s(0)$ is maximum) we find a reasonably constant value of $\rho_s(0)$ (Fig. 1), before of course it dives to zero near $x=0.27$ where $T_c$ is also zero. The temperature dependence is also in good agreement with the weak-coupling $d$-wave temperature dependence. Notably, $\rho_s(T)$ for $x=0.24$ still shows significant deviations from the $d$-wave curve that possibly reflect changes in the electronic structure [8]. In fact the data for $x=0.24$ are in excellent agreement with a weak-coupling $d$-wave calculation for a rectangular Fermi surface [28]. This would not be surprising given the changes in Fermi surface with the rapid crossover from hole-like to electron-like states in that region, as observed by angular resolved photoemission spectroscopy experiments [29]. In the optimal and underdoped regions $\rho_s(0)$ is rapidly suppressed and there is a marked departure of the temperature dependence of $\rho_s(T)$ from the canonical weak coupling curve. Regarding the accuracy of the data we note that each data point represents a total of 16 samples. Given that all our samples were prepared under the same conditions, the size and shape of the grains were essentially the same for all Sr concentrations and the $\lambda_{ab}(0)$ values measured by both the $ac$-susceptibility and $\mu$SR techniques are in excellent agreement, we believe the actual error is significantly lower than the estimates shown in Fig. 1. It is important to note that as for $\lambda_{ab}$ we have also observed a similar behaviour in the $c$-axis component. In Fig. 2 we show our $\lambda_c^{-2}(0)$ data for La-214 and $HgBa_2CuO_{4+\delta}$ (Hg-1201) [12]. Both HTS show a clear drop in $\lambda_c^{-2}(0)$ near $x_{opt}$, with Hg-1201 displaying the sharpest change of all HTS to date, which could be related to the high degree of homogeneity and order in this monolayer cuprate. (The doping level in Hg-1201 was determined by both thermoelectric power and the universal relation $T_c=T_{c,max}[1-82.6(x-0.16)^2]$. [12])

The doping dependence of the superfluid density instructs a well defined direction for further investigations on the next fundamental quantity in the physics of these doped antiferromagnetic insulators, *i.e.*, the magnetic ground state. Appreciating the presence of a competition between several distinct ground states in these frustrated systems one is guided towards the investigation of the low energy dynamics *e.g.*, a spin, charge glassy response. To make a detail comparison we decided to perform a step-by-step investigation. We started from the parent insulator and measured the low energy dynamics by doping in small increments: In La-214 the parent 2D antiferromagnetic insulator $La_2CuO_4$ displays a sharp peak in the magnetic susceptibility at the Néel temperature $T_N$=300K. $T_N$ decreases fast with hole-doping and the transition width broadens (Fig. 3 – upper panel). Experimental evidence from various techniques, and on several HTS, show the emergence with the first added carriers of a second freezing transition $T_F$ as shown in (Fig. 3) for results obtained by μSR. The peak in the spin lattice relaxation rate (Fig. 3 - lower panel) for example is one of the indications for the short range nature of this second freezing.

For $x$>0.02, $T_N$=0 but spin freezing at lower temperature persists. As shown in Fig. 3 (inset in the upper panel) with $x$=0.03 the short range order has been fully exposed, and we can now perform thermodynamic studies to gain further insight into its nature. Here, the low-field susceptibility displays a well-defined cusp, and a thermal hysteresis is observed at $T<T_g$ (Fig. 3 - upper panel), where the material displays scaling, memory effects like traditional glasses and is described by an Edwards-Anderson order parameter [30-32]. Although the signal of this glassy order is somewhat rounded at $T_g$, and the field-cooled curve shows a small but distinct increase with decreasing temperature, as compared to other traditional spin glass systems such as Au-Fe or Cu-Mn, the basic characteristics of the glass order agree very well. (I assign the subscript "g" since on the basis of the similarities of the susceptibility data with those for classic glasses, I believe we have a better reason to call it a glass than something else. I retain the subscript "F" for $x$<0.02 since at this doping region we cannot yet (due to insufficient measurement resolution) confirm the glassy systematics we do for 0.02<$x$<0.05. I note however, the μSR data suggest that the dynamics of the measured short range glassiness are similar to those for $x$>0.02.)

In light of the unconventional superconducting and normal state properties of HTS it is important to probe the possible correlation between the identified glassiness and superconductivity. μSR has been successful in identifying the freezing of electronic moments under the superconducting dome of various HTS [8,11,22-27]. Figure 4 shows the typical time dependence at several temperatures of the zero-field muon asymmetry for the pure bulk superconductor La-214 (*i.e.*, $y=0$) with $x=0.08$. The high-temperature form of the depolarisation is Gaussian and temperature independent, consistent with dipolar interactions between the muons and their near-neighbour nuclear moments. This was verified by applying a 50 Gauss longitudinal field, which completely suppressed the depolarisation. Here the electronic spins in the $CuO_2$ planes fluctuate so fast that they do not affect the muon polarisation. On cooling the fluctuations slow down until they enter the μSR time window and begin to contribute to the relaxation of the polarisation. This is clearly evident in the figure with the crossover in the curvature in the time-decay of the asymmetry. At low enough temperatures the fluctuations freeze out and, typical of other magnetic glassy systems, there is a fast relaxation due to a static distribution of random local fields, followed by a long-time tail with a slower relaxation resulting from remnant dynamical processes within a glassy regime.

To study the doping dependence of the slowing down of the moments towards freezing we determine two characteristic temperatures: (i) The temperature, $T_f$, where the spin correlations first enter the μSR time window *i.e.*, where the muon asymmetry first deviates from Gaussian behaviour and (ii) the temperature, $T_g$, at which these correlations freeze into a glassy-like state thus causing an initial rapid decay in the asymmetry. μSR is sensitive to electronic fluctuations within a time window of $10^{-9}$s to $10^{-6}$s and we may therefore associate $T_f$ and $T_g$, respectively, with these lower and upper thresholds. In general the relaxation data may be fitted to the form $G_z(t)=A_1\exp(-\gamma_1 t)+A_2\exp(-(\gamma_2 t)^\beta)+A_3$ where the first term is the fast relaxation in the glassy state (*i.e.*, at higher temperatures $A_1=0$), the second "stretched-exponential" term is the slower dynamical term and $A_3$ accounts for a small time-independent background arising from muons stopping in the silver backing plate. As in other glassy systems, in the high-temperature Gaussian limit $\beta=2$ [21,33]. Consequently, any departure below $\beta=2.0\pm0.06$ (Fig. 5) is taken as the onset temperature, $T_f$, at which spin fluctuations slow down sufficiently to enter the time scale of the muon probe ($10^{-9}$s). As a further check on the assignment of $T_f$ we fitted the high-temperature data to the full Kubo-Toyabe function $G_z(t)=A_1\exp(-\alpha \hat{t})\exp(-\gamma t)+A_2$. The relaxation rate $\gamma$ is found to rise from zero at the same temperature at which the exponent $\beta$ departs from 2, confirming the entrance of the spin correlations into the experimental time window. At low temperatures the exponent $\beta$ falls rapidly towards the value 0.5 as expected for a glass [33]. We identify $T_g$ as the temperature at which $\beta=0.5\pm0.06$. This "root exponential" form for the relaxation function is a common feature of glasses, and in the present samples the temperature $T_g$ coincided with a maximum in the longitudinal relaxation rate $\gamma$ and the appearance of the fast relaxation. Other methods of analysis may still be possible and a different choice might affect the magnitude of the $T_g$ values but not the trends. Our values for $T_g$ agree with published data obtained by different techniques, where available [22,23,25,34-37].

Let me now make a brief overview of the glassy data discussed above for pure La-214 (*i.e.*, $y=0$). Figure 3 shows the development of a short-range magnetic order below a characteristic temperature $T_F$ with the first added carriers. With increasing doping $T_F$ seems to increase, within error, until $x\sim0.02$ where the $T_N$ vanishes. For $x>0.02$ the freezing temperature, $T_g$, of the short-range magnetic phase gradually decreases with further doping. Notably, this behaviour is common in HTS (pure $YBa_2Cu_3O_{6+\delta}$ (Y-123), $Y_{1-y}Ca_yBa_2Cu_3O_{6.02}$ (Ca:Y-123) and La-214) and has been seen both by neutron scattering and $\mu$SR [22,23,25,35-37].

Furthermore, we find that although the freezing occurs at very low temperatures, $T_g$, low-frequency spin fluctuations enter the experimental time window at significantly higher temperatures, $T_f$. Values of $T_g$ and $T_f$ summarised in Fig. 6 indicate that the glass phase persists beyond $x=0.125$. In fact the onset of the glass phase for $x=0.125$ occurs at a higher temperature than that for $x=0.10$. This may be due to the lock-in of stripe domains [38-40]. I also note that although the doping dependence of the glass transition temperature $T_g$ does not seem to change much with the onset of superconductivity (at $x=0.055$), $T_f$ does. (This observation is not surprising in view of the strong modification of the spin spectrum accompanying the onset of superconductivity and seen *e.g.* in inelastic neutron scattering by the growth of the magnetic resonance [18].) For $x$=0.15 and 0.17, $T_g$ becomes very small (<40mK) and $T_f$ is approximately 8K and 2K, respectively. For $x$≥0.20, there are no changes in the depolarisation function to the lowest temperature measured (40mK) suggesting that here the fluctuations have very short lifetimes, certainly outside of the $\mu$SR time window. Similar observations have been made in $La_{1.6-x}Nd_{0.4}Sr_xCuO_4$ single crystals where no magnetic order was observed for $x$≥0.20 [26].

It is tempting to infer from the absence of dynamic effects ($T_f$=0) that spin fluctuations are no longer present for $x$>$x_{opt}$. Given the frequency limitation of the $\mu$SR technique one can only say that these spins fluctuate with a lifetime <$10^{-9}$s. Also our failure to observe $T_g$ for $x$>0.15 in the La-214 series (Fig. 6) again may simply be due to the limits set by the measurement technique. As shown in the inset of Fig. 6, $T_g$ and $T_f$ scale rather well suggesting a general systematic slowing of the spin fluctuations about $x_{opt}$. This suggests that an observable $T_f$ signals the presence of an observable $T_g$ yet the latter is subject to limitations of the measurement tool. This is a crucial question, which is addressed below in our Zn substitution studies.

Therefore, it is important to check whether $x_{opt}$ is robust with respect to the energy window or is a number associated with the frequency limit of the technique. To this aim we need to either slow down the spins or expand the frequency window of the technique. In our experiment it is easier to do the former and this is the route we have followed. Spectroscopic studies, including inelastic neutron scattering [18,41] and nuclear magnetic resonance [42] experiments, show that Zn substitution has a two-fold impact in slowing spin fluctuations and suppressing long-range magnetic order. This would imply that Zn doping should promote glassy behaviour. Indeed, as depicted in Fig. 7, Zn enhances the muon depolarisation rate at low temperatures and causes an increase in both $T_g$ and $T_f$. The striking result which Fig. 7 summarises is the apparent convergence of both $T_g(x)$ and $T_f(x)$ to zero, for all Zn concentrations, at $x_{opt}$. While this effect is not so obvious for the pure samples (Fig. 6) it is very clear in the two Zn-substituted series. The fact that $T_f(x)$→0 as $x$→$x_{opt}$ for all Zn concentrations suggests that spin correlations within the upper $\mu$SR time threshold of $10^{-9}$s die out beyond $x_{opt}$ leaving only very short-lived fluctuations (or none at all) beyond $x_{opt}$. The fact that $T_f$ and $T_g$ both vanish as $x \to x_{opt}$ implies that the rate of slowing down diverges at $x_{opt}$, in the sense that the characteristic time changes from $10^{-9}$ and $10^{-6}$s in progressively smaller temperature intervals as $x_{opt}$ is approached. Our studies provide experimental evidence showing abrupt changes in the magnetic spectrum at $x_{opt}$ namely, a similarly sudden disappearance of low-frequency spin fluctuations. The observation of simultaneous changes in the magnetic and quasiparticle spectrum is a pivotal result.

Encouraged by the disappearance of glassy fluctuations where the superfluid density is optimal, I now ask whether our results reflect the presence of an underlying quantum critical point separating the glassy non-Fermi liquid and the metallic-Fermi liquid-like regimes of these materials. To address this question we need to gradually increase the amount of disorder to the extent where superconductivity is suppressed fully throughout the phase diagram, and *expose the glassy ground state*. As discussed above these conditions are met by $Zn^{2+}$ doping. Figure 8 depicts characteristic data for $La_{2-x}Sr_xCu_{1-y}Zn_yO_4$ ($y = 0.05$) with the normal state exposed ($T_c=0$) across the $T$-$x$ phase diagram.

It is clear from the data that changes in the magnetic ground state of La-214 occur at $x=x_{opt}$. (For details to this systematic investigation the reader is refereed to the original article [43].) The main contribution of this study is that superconductivity has been fully suppressed and therefore any masked or hidden short range magnetism is now exposed. Values of $T_g$ summarised in Fig. 8 indicate a gradual decrease of the onset of the spin glass phase with doping. The exception again is the $x=0.12$ sample for which $T_g$ is higher than for $x=0.10$. We note the positive curvature of $T_g(x)$, exactly discussed earlier for pure, 1 and 2% Zn doped La-214 and expected in quasi-2D systems like the HTS. The increase in $T_g$ in the 1/8 region has been previously discussed in terms of stripe domains. The effect of the ordered phase is also evident for $x=0.14$, which from the trend of $T_g(x)$ indicates the latter is higher than expected for this doping. In fact the presence of a static stripe component for $x=0.14$ can be seen in neutron scattering experiments [15]. The glass regime vanishes at $x_{opt}$ and for $x \geq x_{opt}$ we did not observe even an onset ($T_f$) of fluctuations slowing down sufficiently to enter the frequency scale of the muon probe (Fig. 8). Field dependent studies enabled us estimate the internal local field, $B_{local}$, sensed by the implanted muons [43]. Figure 8 includes the doping dependence near $x_{opt}$ of $B_{local}$, measured at $T$=0.03K. This exhibits behaviour similar to $T_g$ and $T_f$ and disappears at precisely $x_{opt}$.

The values for $T_g(x)$ for $y$=0.05 are lower than those found in the studies for $y$=0.02 even though $T_g(x)$ was found to increase systematically for $x$=0, 0.01 and 0.02 (Fig. 7). This occurs because while Zn slows the spin fluctuations, it also dilutes the spins and at high concentration we see the latter effect. Nevertheless, zinc substitution has undoubtedly exposed the ground state, however distorted it might be, in the samples studied here and allowed us to identify the precise location ($x_{opt}$) at which the glass phase disappears. Based also on the systematic tendency of $T_g$ and $T_f$ to vanish at $x_{opt}$ it is unlikely $y$=0.05 has suppressed short-range magnetic order only for $x \geq x_{opt}$. *Therefore, we can safely confirm our earlier indications that the magnetic ground state of HTS changes character at $x_{opt}$.*

The similar doping dependence of $T_g$ for pure and up to 5% Zn doped samples indicates that regardless of a sample being pure, Zn-doped, superconducting or not we obtain a universal behaviour: A set of glassy phase transitions, enhanced near $x$=1/8, and ending at a specific doping $x$=$x_{opt}$ at $T$=0, supporting the presence of a quantum glass transition. Let me pause for a moment to explain in brief what I mean by a glassy phase transition. As discussed earlier, for a sample with $x$=0.03 where the short range order has been fully exposed and superconductivity is absent, we were able to perform thermodynamic studies to gain further insight into the nature of the apparent glassy phase. We as well as other colleagues [30-32,35], found the low-field susceptibility has a well-defined cusp and a thermal hysteresis at $T<T_g$ (Fig. 3 - upper panel), where the material displays scaling, memory effects like traditional glasses and is described by an Edwards-Anderson order parameter [30-32,35]. Although close enough to the transition the system behaves classically, the classical region will shrink as the temperature approaches zero. Therefore, a way to reach the quantum regime would be through the destruction of the glass phase by tuning the ground state at $T$=0 using carrier concentration as a quantum tuning parameter. As discussed above, we did exactly that and found that indeed a large enough concentration of mobile carriers eventually destroyed the glass order, even at $T$=0. The phase transition at $T$=0 would therefore be quantum mechanical.

Since the observation $T_{g,f}(x)=0$ and $B_{local}=0$, at $x \geq x_{opt}$ for all Zn concentrations, our results indicate that the identified glassy state and associated low frequency fluctuations disappear at $x_{opt}$ and the results for lower or no Zn dopings to this effect were not masked by either the frequency window of the technique. Bearing in mind the fact that $T_f$ and $T_g\_0$ at $x_{opt}$ so that the rate of slowing down actually diverges at this point, the present results indicate the existence of a quantum glass transition at $x_{opt}$. Because, (based on extrapolation from the aforementioned magnetisation studies for $x<0.05$ [14,30-32,35], and the systematic trends of the $\mu$SR spectra across the many different samples we have so-far investigated) the quantum glass transition has an associated order parameter and symmetry breaking due to the frozen spins at $T=0$, we may interpret the glass transition at $T=0$ as a quantum critical point.

Therefore, in the absence of evidence for long-range order in the normal state but in the presence of a glassy phase transition, the present observations indicate the existence of a quantum phase transition at $x_{opt}$. In fact, glass order is an attractive candidate for a quantum critical point in electronically disordered systems - impurities break the translational symmetry, prohibiting a quantum critical point in the presence of randomness in these quasi 2D systems. Furthermore, because even with a large amount of disorder which is enough to even suppress high temperature superconductivity across the whole phase diagram the doping dependence of $T_g$ remains essentially the same as in the pure system, the observed glassiness must be predominantly self-generated. That is, it emerges by doping the parent low dimensional antiferromagnetic insulator with carriers.

I would like to note the present results are not peculiar just to La-214. La-214 is sometimes said to be atypical of HTS because of its apparently high degree of inhomogeneity. Yet we have observed similar behaviour in the $Bi_{2.1}Sr_{1.9}Ca_{1-x}Y_xCu_2O_{8+\delta}$ (Bi-2212) system [11]. We note further that the doping dependence of $T_g$ seen here has also been found in pure Y-123 [37] and in Ca:Y-123 up to 0.09 holes per planar copper atom [25] suggesting that the behaviour discussed here is generic and common to all high-$T_c$ materials. (Unfortunately, $\mu$SR investigations on the $YBa_2Cu_3O_{6+\delta}$ (Y:123) family for oxygen concentrations $\delta>0.6$ – corresponding to approximately $x>0.10$, is not possible to test the disappearance of glassiness where superconductivity is optimal, $\delta=0.95$. For $\delta>0.6$, the majority of the muons reside on the now filled Cu-O chains, hence probing the magnetic response of the glassy $CuO_2$ planes is no longer possible. Recent nuclear quadrupole resonance experiments [44] however, provided evidence in support of the presence of spin/charge inhomogeneity to dopings higher than previously reported by $\mu$SR [25,35]. I believe it is a matter of time before we are in a position to probe the low energy dynamics of this system too up to the slightly overdoped region.) Notably, values of $T_g$ for the Y-123 and Bi-2212 families, believed to pose less disorder than La-214, are in fact twice the values of the latter. At the same time Y-123 is generally regarded as a much cleaner system than Bi-2212. Similarly, $T_g$ for pure Y-123 is about the same as $T_g$ for Ca:Y-123 in spite of the substantial disorder in the latter system [25,35]. Therefore, the emergence of a glassy state in these quasi-2D antiferromagnetic insulators is generic and a property of the parent insulator doped with carriers. Unlike the La-214 family, spanning with doping the whole phase diagram with doping and exposing the ground state without causing macroscopic phase separation is not yet possible in other HTS. It would therefore take time before a systematic investigation as the one presented here for La-214 can be made to test the proposed universality in detail in the other HTS families.

Our detailed *ac*-susceptibility data for the penetration depth served the important role in identifying directly the interplay between magnetism and superconductivity: The results revealed a close relationship between the onset of slow spin fluctuations (and glass behaviour at low temperature) and the observed sharp reduction in superfluid density at $x_{opt}$. This shows, in particular, that the onset of short-range magnetic correlations coincides with an abrupt change in the superconducting ground state. At higher dopings the ground state becomes metallic and homogeneous, with no evidence for glassiness. These results provide unambiguous evidence of a sharp change in ground state properties at a specific doping and the emergence of vanishing temperature scales as this point is approached - just as one expects at a quantum critical point. Moreover, it is in this doping region where the Fermi surface topology of La-214 changes from hole-like to electron-like.

**Summary**

Following the changes in the superconducting ground state as probed by measurements of the superfluid density, we identified signatures for a quantum critical point intercepted by glass order where superconductivity is strongest. One may be optimistic and envisage a scenario, where further to the magnetic interactions associated with the magnetic instability, slow density fluctuations, due to the spontaneous tendencies toward glassy formation may induce an additional attractive quasiparticle interaction.

**Acknowledgements**

I am grateful to my collaborators C.W. Chu, J.R. Cooper, V. Dobrosavljevic, P.P. Edwards, A. Hillier, M. Kodama, T. Nishizaki, A. Petrovic, B.D. Rainford, T. Sasagawa, C.A. Scott, J.L. Tallon and T. Xiang, and The Royal Society (London), Trinity College (Cambridge), and the Rutherford Appleton Laboratory for their valuable contributions.

# References

1. D. Shoenberg *Superconductivity* (Cambridge University Press, Cambridge, 1954), p.164.

2. C. Panagopoulos *et al.*, Phys. Rev. B (*Rapid Comm.*) **53**, R2999 (1996).

3. C. Panagopoulos *et al.*, Phys. Rev. B (*Rapid Comm.*) **54**, R12721 (1996).

4. T. Xiang, C. Panagopoulos and J.R. Cooper, Int. J. Mod. Physics B **12**, 1007 (1998).

5. C. Panagopoulos, J.R. Cooper and T. Xiang, Phys. Rev. B **57**, 13422 (1998).

6. C. Panagopoulos and T. Xiang, Phys. Rev. Lett. **81**, 2336 (1998).

7. C. Panagopoulos, J.L. Tallon and T. Xiang, Phys. Rev. B (*Rapid Comm.*) **59**, R6635 (1999).

8. C. Panagopoulos *et al.*, Phys. Rev. B **60**, 14617 (1999).

9. C. Panagopoulos *et al.*, Phys. Rev. B (*Rapid Comm.*) **61**, R3808 (2000).

10. T. Xiang and C. Panagopoulos, Phys. Rev. B **61**, 6343 (2000).

11. C. Panagopoulos *et al.*, Phys. Rev. B **66**, 064501 (2002).

12. C. Panagopoulos *et al.*, Phys. Rev. B (*Rapid Comm.*) **67**, R220502 (2003).

13. S. Sachdev, Rev. Mod. Phys. **75**, 913 (2003).

14. C. Panagopoulos and V. Dobrosavljevic, Phys. Rev. B **72**, 014536 (2005)

15. S. A. Kivelson *et al.*, Rev. Mod. Phys. **75**, 1201 (2003).

16. N.D. Mathur *et al.*, Nature (London) **394**, 39 (1998).

17. P.G. Radaelli *et al.*, Phys. Rev. *B* **49**, 4163 (1994).

18. M.A. Kastner, R.J. Birgeneau, G. Shirane, Y. Endoh, Rev. Mod. Physics **70**, 897 (1998).

19. C. Panagopoulos *et al.*, Physica C **269**, 157 (1996).

20. J. Chrosch, C. Panagopoulos, N. Athanassopoulou, J. R. Cooper and E.K.H. Salje, Physica C **265**, 233 (1996).

21. Y.J. Uemura *et al.*, Phys. Rev. B **31**, 546 (1985).

22. D. R. Harshman *et al.*, Phys. Rev. B **38**, 852 (1988).

23. J. I. Budnick *et al.,* Europhys. Lett. **5,** 65 (1988).

24. R.F. Kiefl *et al.,* Phys. Rev. Lett. **63**, 2136 (1989).

25. Ch. Niedermayer *et al.,* Phys. Rev. Lett. **80**, 3843 (1998).

26. B. Nachumi *et al.,* Phys. Rev. B **58**, 876 (1998).

27. A. Kanigel, A. Keren, Y. Eckstein, A. Knizhnik, J.S. Lord and A. Amato, Phys. Rev. Lett. **88**, 137003 (2002).

28. T. Xiang and J.M. Wheatley, Phys. Rev. Lett. **77**, 4632 (1996).

29. A. Ino *et al.,* J. Phys. Soc. Jpn. **68**, 1496 (1999).

30. F. C. Chou, N.R. Belk, M.A. Kastner, R.J. Birgeneau and A. Aharony, Phys. Rev. Lett. **75,** 2204 (1995).

31. A. Lavrov *et al.,* Phys Rev Lett. **87**, 017007 (2001).

32. S.F. Edwards and P.W. Anderson, J. Phys. F: Metal Phys. **5**, 965 (1975); *ibid* **6**, 1927 (1976).

33. R. Cywinski and B.D. Rainford, Hyperfine Interact. **85**, 215 (1994).

34. P.M. Singer and T. Imai, Phys Rev Lett. **88**, 187601 (2002).

35. S. Wakimoto, S. Ueki, Y. Endoh and K. Yamada, Phys. Rev. B **62**, 3547 (2000).

36. M. Matsuda *et al.,* Phys. Rev. B **65**, 134515 (2002).

37. S. Sanna, G. Allodi, G. Concas, A. H. Hillier, and R. D. Renzi, Phys. Rev. Lett. **93**, 207001 (2004).

38. J.M. Tranquada *et al,.* Phys. Rev. Lett. **78**, 338 (1997).

39. J. Zaanen, J. Phys. Chem. Solids **59**, 1769 (1998).

40. C.M. Smith, A.H. Castro Neto and A.V. Balatsky, Phys. Rev. Lett. **87**, 177010 (2001).

41. H. Kimura *et al.,* Phys. Rev. B **59**, 6517 (1999).

42. M.-H. Julien *et al.,* Phys. Rev. Lett. **84**, 3422 (2000).

43. C. Panagopoulos *et al.,* Phys. Rev. B **69**, 144510 (2004).

44. R. Ofer, S. Levy, A. Kanigel and A. Keren, Phys. Rev. B **73**, 012503 (2006).

**Figure captions**

Fig. 1. The upper panel depicts the doping dependence of the absolute value of the superfluid density for grain-aligned $La_{2-x}Sr_xCuO_4$. The lower panel shows the temperature dependence of the normalised superfluid density for the same material at various dopings compared with the weak-coupling BCS theory (solid line) for a *d*-wave superconductor.

Fig. 2. Doping dependence of the inverse square of the absolute value of the *c*-axis penetration depth for two monolayer HTS, for $La_{2-x}Sr_xCuO_4$ (La-214) and $HgBa_2CuO_{4+\delta}$ (Hg-1201).

Fig. 3. The upper panel depicts the doping dependence of the Néel temperature $T_N$, the second freezing $T_F$, and glass temperature $T_g$ for $La_{2-x}Sr_xCuO_4$ (La-214). On the top right of the panel the plot depicts the temperature dependence of the susceptibility for *x*=0 and 0.01 single crystals with *H//c*. The associated inset is data for *x*=0.03 showing the transformation of the material to a glass (closed and open circles, are results for zero-field-cooled and field-cooled experiments, respectively). The lower panel depicts a peak in the spin lattice relaxation $1/T_1$ data obtained by zero field *μ*SR for *x*=0.01.

Fig. 4. Zero-field *μ*SR spectra of $La_{2-x}Sr_xCuO_4$ for *x*=0.08 measured at different temperatures. The solid lines are the fits discussed in the text.

Fig. 5. Typical temperature dependence of the exponent *β* of $La_{2-x}Sr_xCuO_4$ for *x*=0.08-0.17. The left-hand panel shows linear plots with an arrow showing $T_f$ for *x*=0.10, whereas as in the right-hand panel we show semi-log plots with the arrow indicating a $T_g$.

Fig. 6. Doping dependence of $T_g$ (closed circles), $T_f$ (open circles) and $T_c$ (crosses) of of $La_{2-x}Sr_xCuO_4$. The inset is a semi-log plot of $T_g$ (mulitplied by 12) and $T_f$ as a function of doping.

Fig. 7. Doping dependence of $T_g$ (closed symbols) and $T_f$ (open symbols) of $La_{2-x}Sr_xCu_{1-y}Zn_yO_4$; $y=0$ (circles), $y=0.01$ (triangles) and $y=0.02$ (diamonds). The values of $T_c$ are shown as crosses for all values of $y$ (indicated in the figure).

Fig. 8. The doping dependence of the temperature $T_g$ (closed circles), below which the spin fluctuations freeze out into a glass, for $La_{2-x}Sr_xCu_{1-y}Zn_yO_4$ ($y= 0.05$; $x=0.10$-$0.22$). The isnet depicts data for $T_f$ and $B_{local}$ near $x_{opt}$.

**Fig. 1**

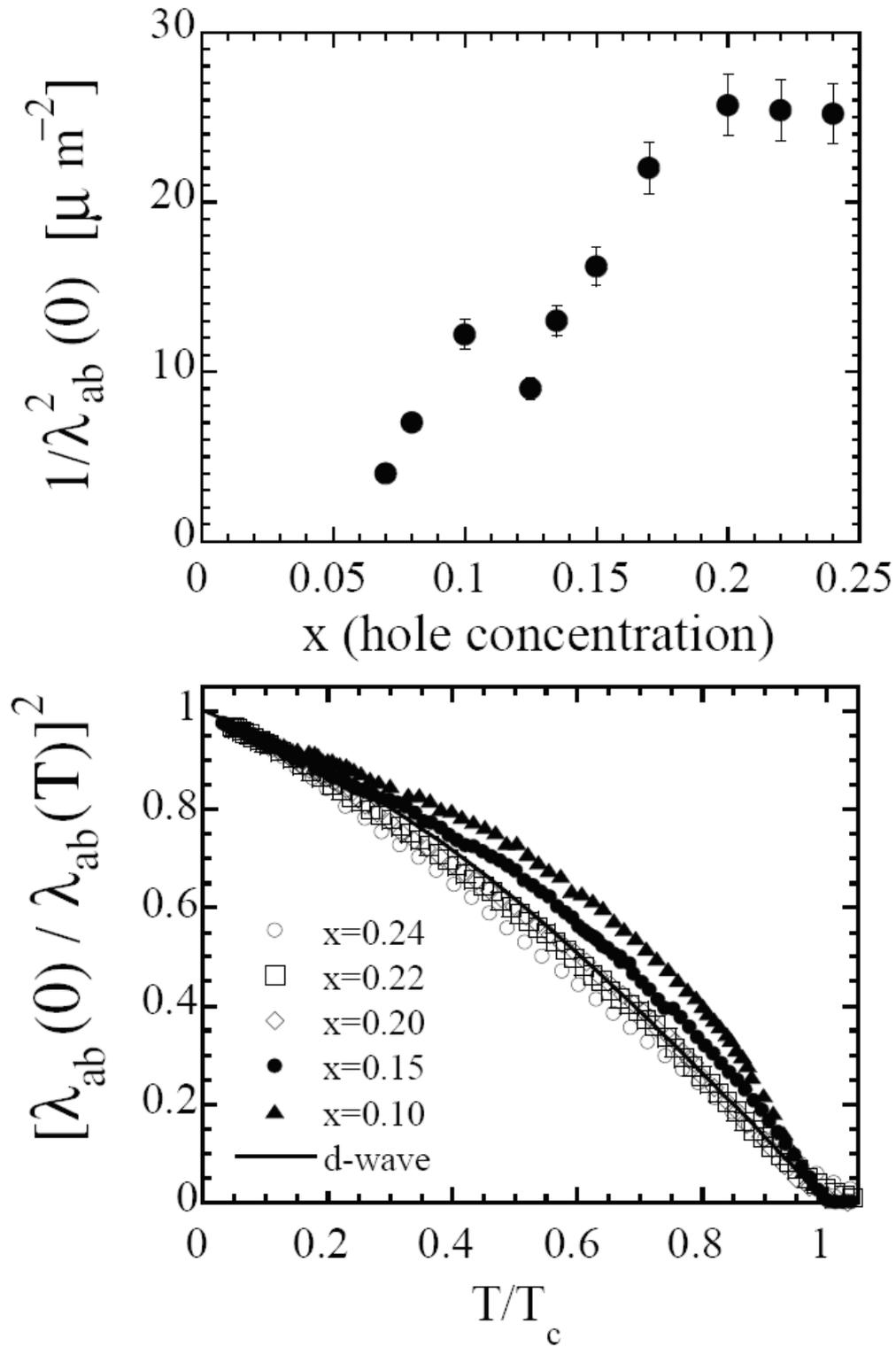

**Fig. 2**

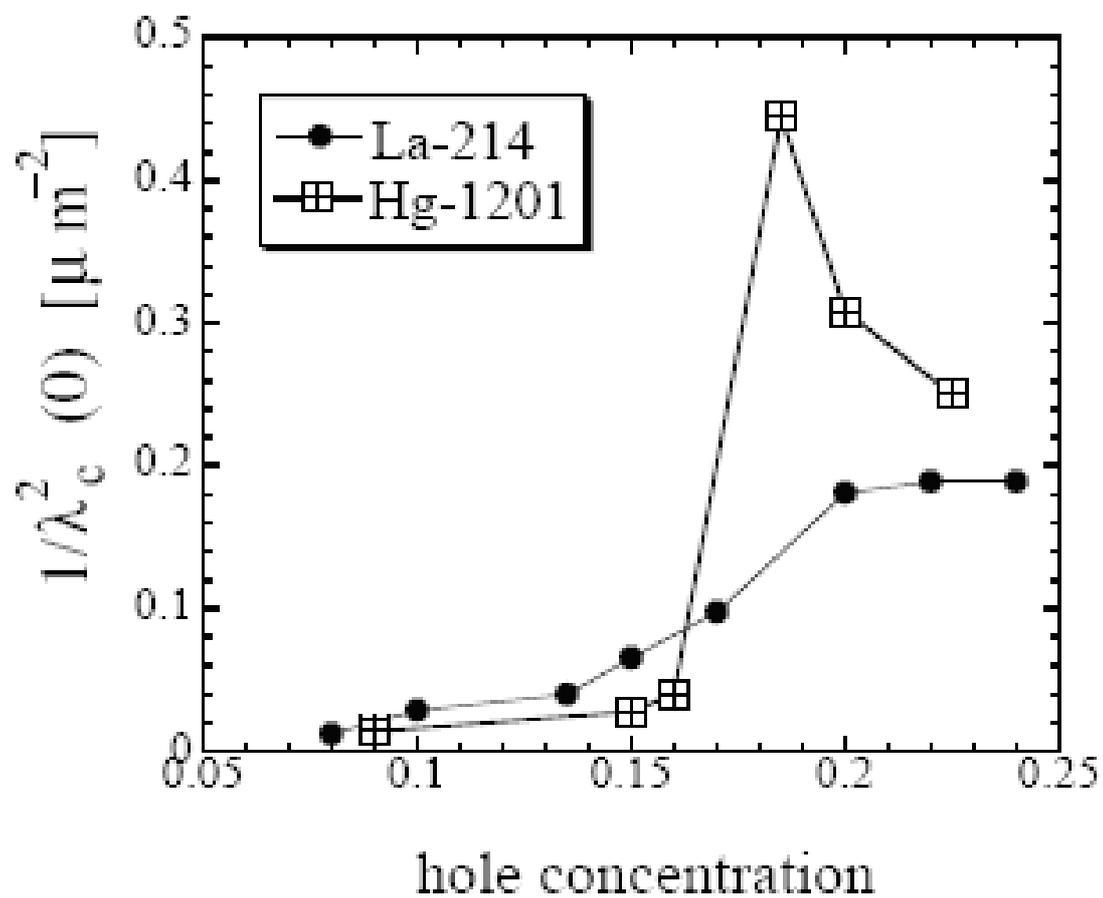

**Fig. 3**

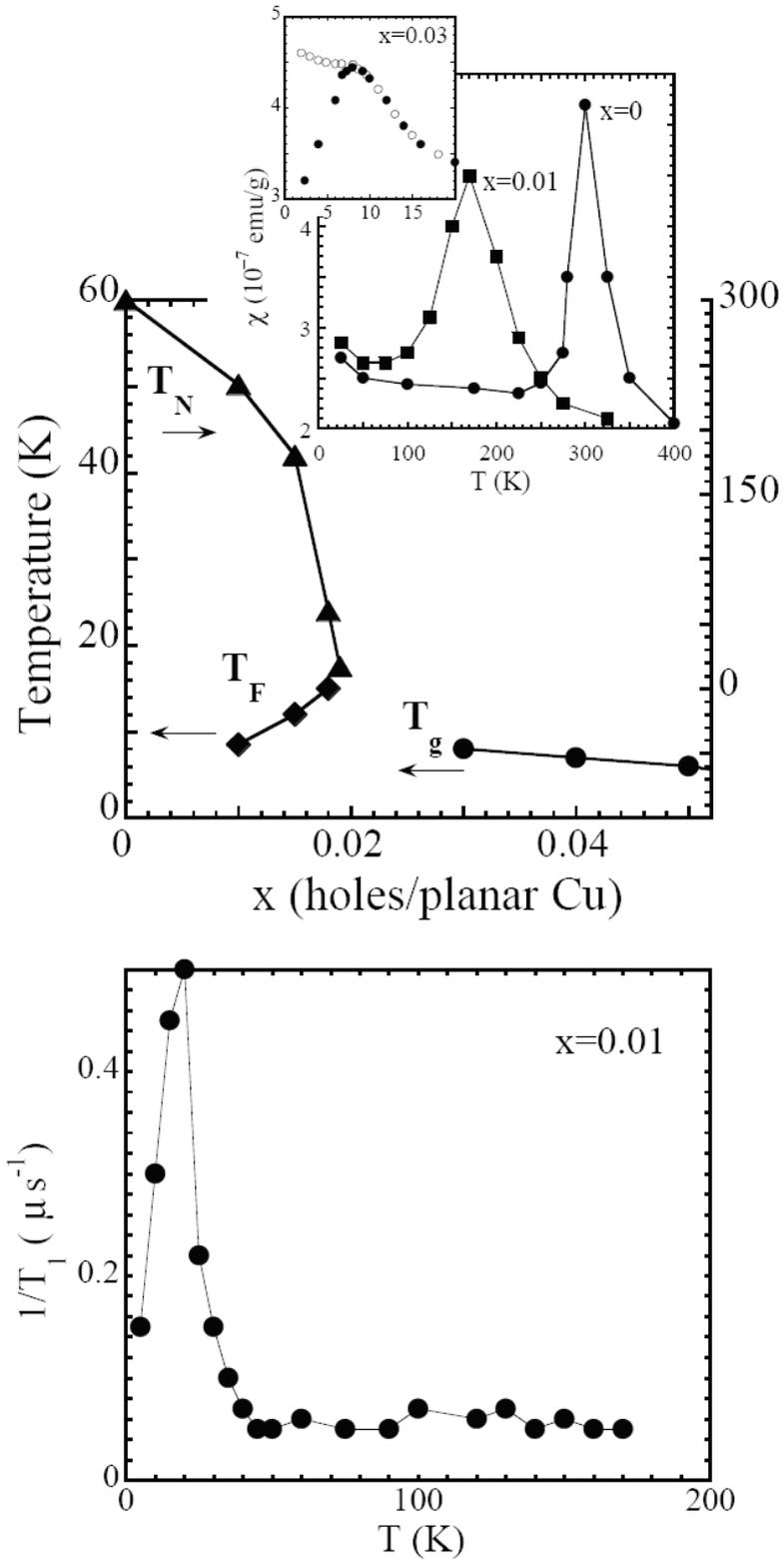

**Fig. 4**

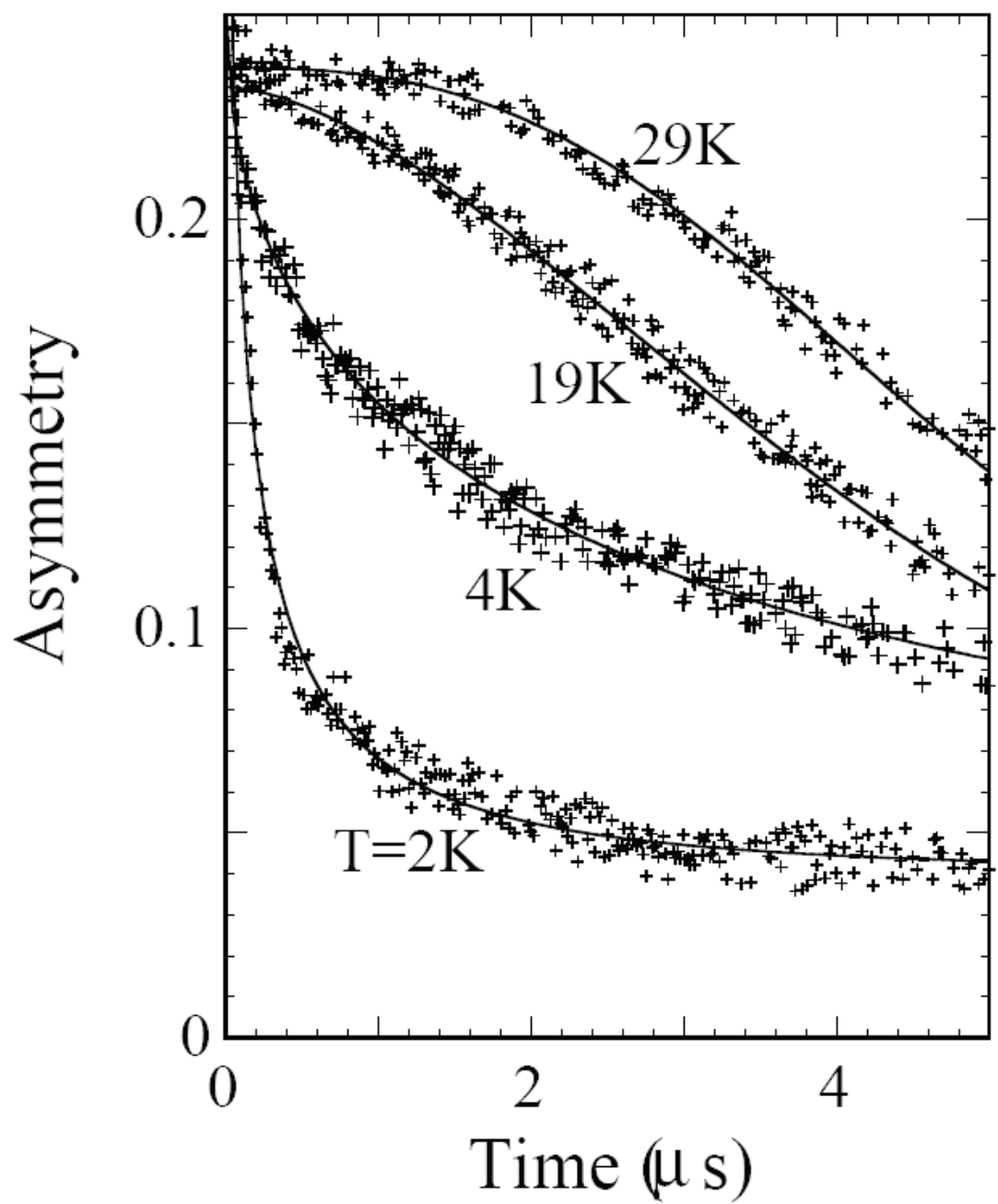

**Fig. 5**

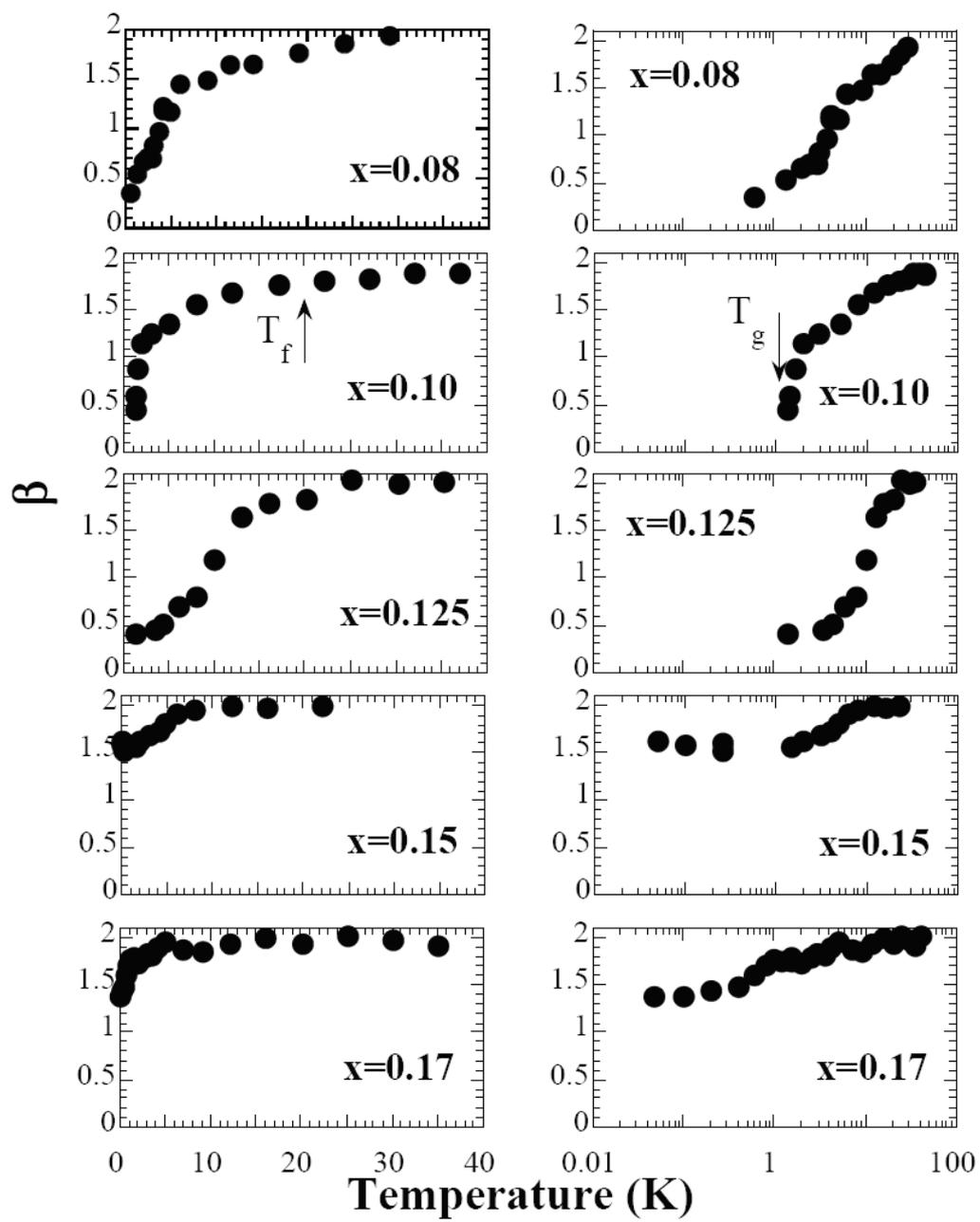

**Fig. 6**

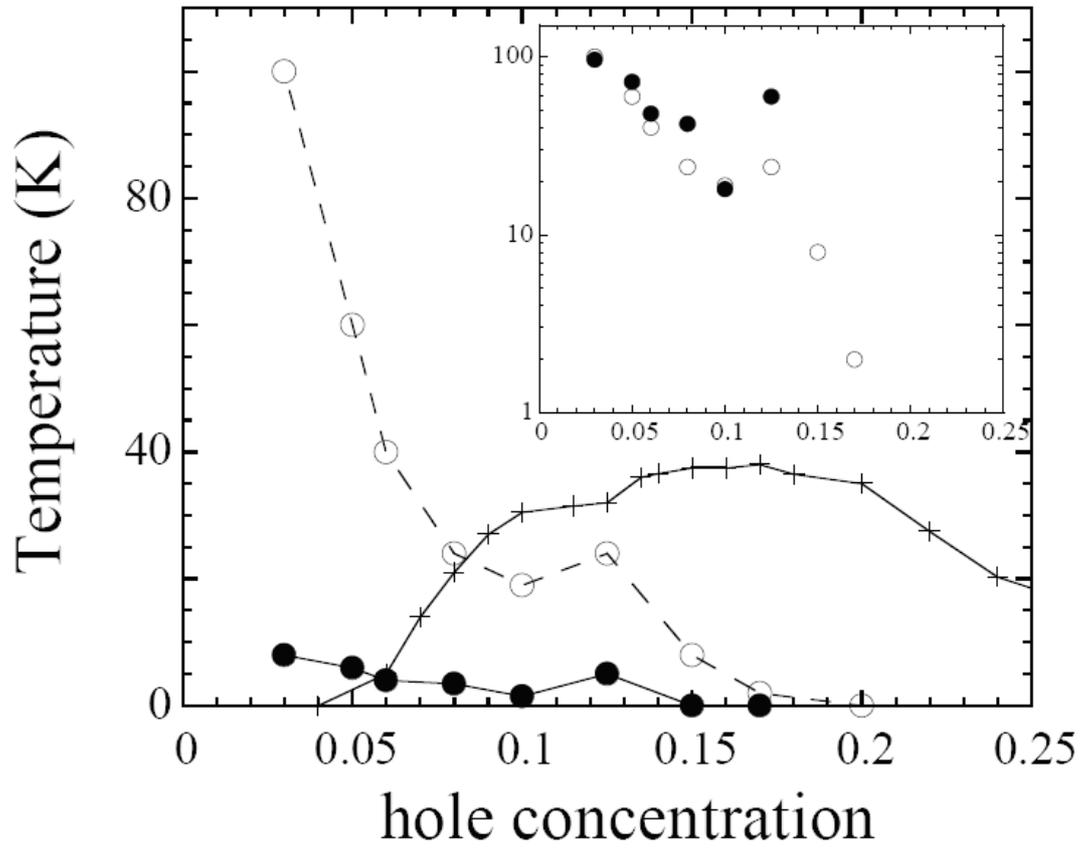

**Fig. 7**

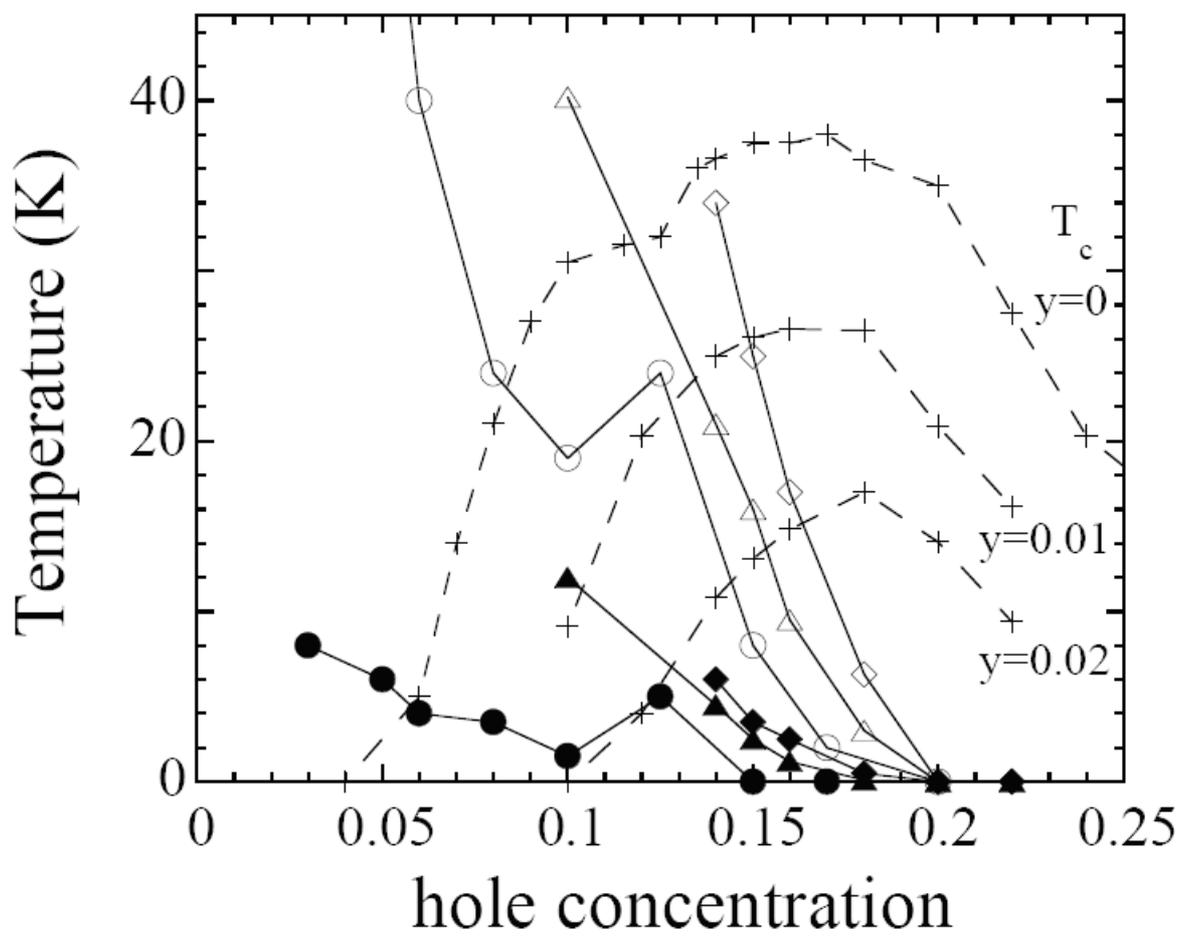

**Fig. 8**

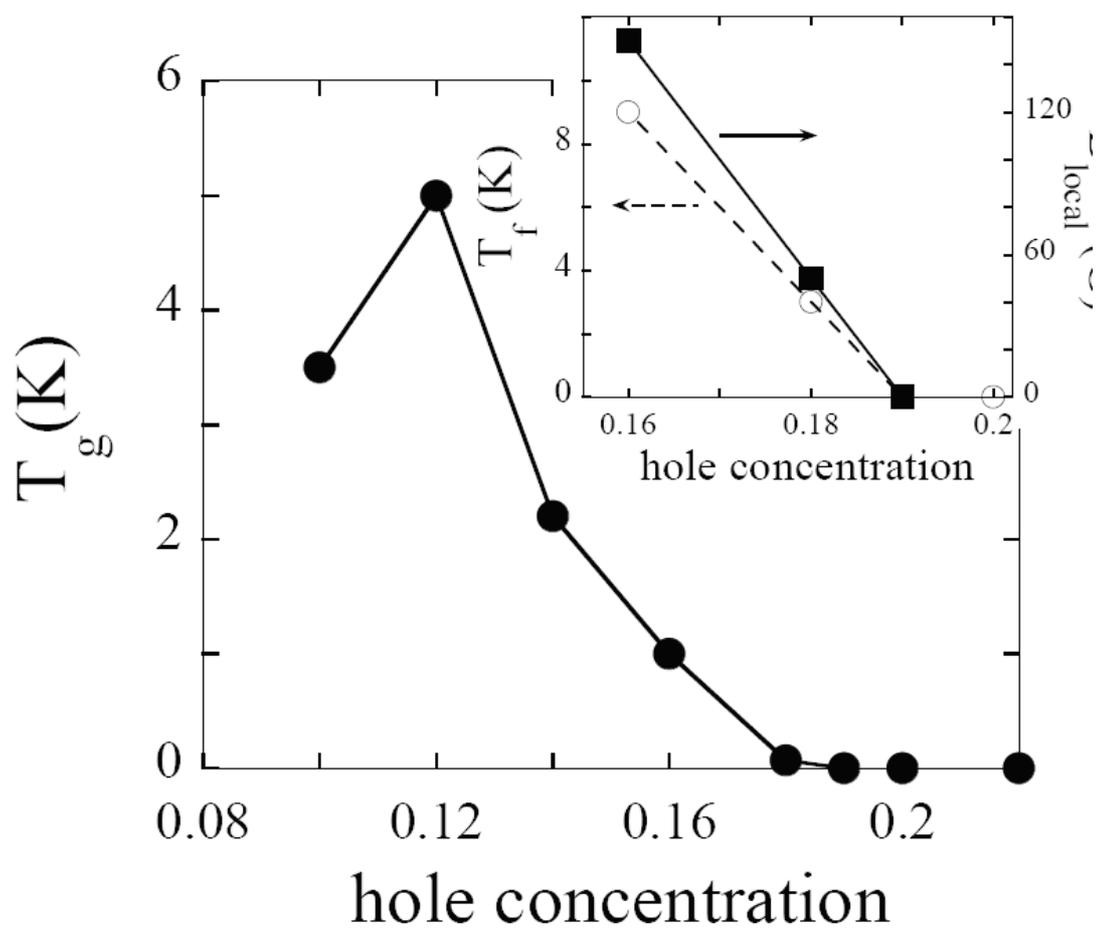